\documentclass[aps,pre,reprint,groupedaddress,amsmath,amssymb]{revtex4-1}

\usepackage{bm}
\usepackage{graphicx}
\usepackage{color}

\begin{document}

%\preprint{}

\title{Phase reduction of a limit cycle oscillator perturbed by a strong amplitude-modulated high-frequency force}


\author{Kestutis Pyragas and Viktor Novi\v{c}enko}
%\thanks{}
\affiliation{Center for Physical Sciences and Technology, A. Go\v{s}tauto 11, LT-01108 Vilnius, Lithuania}


\date{\today}

\begin{abstract}
The phase reduction method for a limit cycle oscillator subjected to a strong amplitude-modulated high-frequency force is developed. An equation for the phase dynamics is derived by introducing a new, effective phase response curve. We show that if the effective phase response curve  is everywhere positive (negative), then an entrainment of the oscillator to an envelope frequency is possible only when this frequency is higher (lower) than the natural frequency of the oscillator. Also, by using the Pontryagin maximum principle, we have derived an optimal waveform of the perturbation that ensures an entrainment of the oscillator with minimal power. The theoretical results are demonstrated with the Stuart-Landau oscillator and  model neurons.
\end{abstract}

%\keywords{}

\maketitle


\section{Introduction} Self-sustained oscillations  are of great interest for the physical, chemical and biological sciences \cite{kura03, winf01, pikov01, izhi07}. The oscillations appear in nonlinear dissipative systems and are typically modeled by limit cycle oscillators. The phase reduction method \cite{kura03, winf01} provides a fundamental theoretical technique to approximate high-dimensional dynamics of limit cycle oscillators with a single \emph{phase variable} that characterizes timing of oscillation. This method has been widely and successfully applied to weakly coupled oscillators as well as an oscillator subjected to a weak external force. Various waveform optimization problems have been solved in the framework of this approach to improve entrainment properties of forced spiking neurons \cite{moeh2006, *harada10, *dasa11, *zlotnik2013, *Dasanayake2015}.

In recent years, several extensions of the phase reduction theory have been elaborated. The theory has been successfully adapted to stochastic~\cite{yoshi08, *teram09, *gold10}, delay-induced~\cite{physd12, *kot12}, and collective ~\cite{kawam08, *kawam11} oscillators. Despite the fact that the conventional phase reduction theory deals only with weak perturbations,  Kurebayashi \textit{et al.}~\cite{kureb13} have recently demonstrated that this fundamental limitation can be overcome in some cases. They extended the phase reduction method for a special class of \emph{strong} perturbations that can be decomposed into a strong slowly varying component and remaining weak fluctuations.

In this paper, we extend the phase reduction theory for another class of strong perturbations. We consider a limit cycle oscillator driven by a \emph{strong amplitude-modulated high-frequency} (AMHF) force [e.g., proportional to  $\sin(\Omega t) \sin(\omega t)$] with a carrier frequency $\omega$ considerably greater than the natural frequency $\Omega_0$ of the oscillator and an envelope frequency $\Omega$ comparable to $\Omega_0$. We  derive an equation for the phase dynamics using a combination of an averaging method \cite{sand07, burd07} and the conventional phase reduction approach.

The AMHF perturbations are widely used in neuroscience for controlling synchronization processes in neuronal networks \cite{brum75, *harnack2004, *tass08, tass2003, *tass2003_b, *tass2012, *tass2014}. An innovative therapeutic procedure clinically approved for the treatment of Parkinson's disease, essential tremor and dystonia  is a deep brain stimulation \cite{benabid1991, *marks2005}, in which electrical pulses are applied to inhibit pathological synchrony among the neurons \cite{Lozano2004, *Uhlhaas2006, *Hammond2006}. One of stimulation techniques, referred to as a coordinated reset neuromodulation \cite{tass2003, *tass2003_b, *tass2012, *tass2014}, desynchronizes a neural population via brief, high-frequency pulse trains, which are periodically delivered at different sites of the population (subpopulations) with shifted phases. The need for the mild stimulation protocols raises a challenging problem: how to reset a phase of the subpopulation with the least invasiveness. Regarding this question, we formulate an AMHF \emph{waveform optimization} problem to attain an entrainment of a limit cycle oscillator with minimal power. We solve the problem by employing our developed phase reduction method and the Pontryagin maximum principle \cite{pontryagin1962}.

The paper is organized as follows. In Sec. \ref{sec2} we present our phase reduction theory and demonstrate its validity using two specific examples, namely, the Stuart-Landau oscillator and the Morris-Lecar \cite{morris81} model neuron. Section  \ref{sec3} is devoted to the waveform optimization problem. To numerically demonstrate this theory we use  the FitzHugh-Nagumo \cite{Fitzhugh61, *Nag62} model neuron. A summary is presented in Sec. \ref{sec4}.

\section{\label{sec2} Phase reduction theory}  Let us consider an unperturbed dynamical system $\dot{\mathbf{x}} = \mathbf{f}\left(\mathbf{x} \right)$ with $\mathbf{x}(t)\in \mathbb{R}^n$ and $\mathbf{f}: \mathbb{R}^n \rightarrow \mathbb{R}^n$ and assume that it has a stable $T_0$-periodic limit cycle solution $\mathbf{x}(t)=\bm{\xi}(t)=\bm{\xi}(t+T_0)$. We seek to develop a phase reduction theory for the oscillator driven by a strong AMHF perturbation
\begin{equation}
\dot{\mathbf{x}} = \mathbf{f}\left(\mathbf{x} \right)+ \mathbf{K} \bm{\psi}(\Omega t) \varphi(\omega t),
\label{main}
\end{equation}
where $\mathbf{K}=\textrm{diag}[K_1,K_2,\ldots,K_n]$ is a diagonal coupling matrix, $\bm{\psi}(\Omega t)=[\psi_1(\Omega t),\ldots,\psi_n(\Omega t)]^T$ is an $n$-dimensional envelope vector and $\varphi(\omega t)$ is a scalar high-frequency (HF) carrier signal. The both functions $\bm{\psi}(s)$ and $\varphi(s)$  are $2\pi$-periodic with respect to $s$. We analyze an entrainment of the oscillator to the envelope frequency $\Omega$ assuming that it is close to the frequency $\Omega_0=2\pi/T_0$ of the limit cycle, while   $\omega \gg \Omega_0$. The ratio $\omega/\Omega$ is assumed to be an integer number so that the product $\bm{\psi}(\Omega t) \varphi(\omega t)$ is a periodic function with the same period $T=2\pi/\Omega$ as the envelope. For the HF function $\varphi(\omega t)$, we require the zero average, $\int_0^{2\pi}\varphi(s)ds=0$. In terms of neurostimulation, this constraint represents a charge-balanced requirement, which is clinically mandatory to avoid tissue damage \cite{brum75, *harnack2004, *tass08}. In addition, we assume without loss of generality that the maximum of the function $\varphi(s)$ is equal to $1$ and the minimum is not bellow $-1$, moreover each component $\psi_j(s)$ is in the interval $[-1,1]$ and at least one time during the period reaches one of the boundary.

We are interested in the case when the components of the coupling matrix $\mathbf{K}$ are not small in comparison to the corresponding components of the vector field $\mathbf{f}\left(\mathbf{x} \right)$ so that the conventional phase reduction approach does not apply. Here we develop a modified approach that allows us to derive a phase equation for the system (\ref{main})  in the limit of high frequency $\omega \to \infty$ even when the perturbation is large. Considering this limit it is convenient to scale the coupling matrix as $\mathbf{K}=\omega \mathbf{A}$ with the components of the matrix  $\mathbf{A}=\textrm{diag}[A_1,A_2,...,A_n]$ being independent of $\omega$, i.e., we replace the set of independent parameters $(\omega, \mathbf{K})$ by the set of independent parameters $(\omega, \mathbf{A})$.   Due to the one-to-one relation between the above parameter spaces, the solution found in the space of the parameters $(\omega, \mathbf{A})$ can be uniquely transformed into the original space of the parameters $(\omega, \mathbf{K})$. A motivation for such a transformation of the parameters can be found in the Appendix of Ref. \cite{rat12}.  Let us introduce a particular antiderivative of the HF function as:
\begin{equation}
\Phi(s) = \Phi_1(s)-\left\langle \Phi_1 \right\rangle,
\label{antider}
\end{equation}
where $\Phi_1(s)=\int_0^s \varphi(s')ds'$ and the angle brackets $\left\langle \Phi_1 \right\rangle=(1/2\pi)\int_0^{2\pi}\Phi_1(s)ds$ denote the averaging of a function over its period. The function $\Phi(s)$ has the properties $d\Phi(s)/ds=\varphi(s)$, $\Phi(s+2\pi)=\Phi(s)$ and $\left\langle \Phi \right\rangle=0$. Using this function, we change the variable   $\mathbf{y}(t)=\mathbf{x}(t)-\Phi(\omega t)\mathbf{A}\bm{ \psi}(\Omega t)$ of the system (\ref{main}) and rewrite it as:
\begin{equation}
\dot{\mathbf{y}} = \mathbf{f}\left(\mathbf{y}+\Phi(\omega t)\mathbf{A}\bm{\psi}(\Omega t) \right)- \Phi(\omega t)\mathbf{A} \frac{d}{dt}\bm{\psi}(\Omega t).
\label{main_new_var}
\end{equation}
By introducing an envelope phase variable $\alpha = \Omega t$ and the ``fast'' time variable $\tau=\omega t$, system (\ref{main_new_var}) can be transformed into the standard form of equations as typically used by the method of averaging \cite{sand07}:
\begin{subequations}
\label{slow}
\begin{eqnarray}
\omega\frac{d\mathbf{y}}{d\tau} &=&  \mathbf{f}\left(\mathbf{y}+ \Phi(\tau)\mathbf{A}\bm{\psi}(\alpha) \right) - \Phi(\tau)\mathbf{A} \Omega \frac{d \bm{\psi}(\alpha)}{d \alpha},  \label{slow_main}\\
\omega \frac{d \alpha}{d\tau}  &=& \Omega \label{slow_phase}.
\end{eqnarray}
\end{subequations}
Due to the large  factor $\omega$ƒ in the left hand side (l.h.s.)
of the Eqs. (\ref{slow}), the variables $\mathbf{y}$ and $\alpha$
vary slowly while the periodic function $\Phi(\tau)$ in
the right hand side (r.h.s.) oscillates fast. According to the method of averaging \cite{sand07}, an approximate solution of system (\ref{slow}) can be obtained by averaging the r.h.s. of the system over fast oscillations. Specifically, let us denote the variables of the averaged system as $\bar{\mathbf{y}}$ and $\bar{\alpha}$. They satisfy the equations
\begin{subequations}
\label{slow1}
\begin{eqnarray}
\omega\frac{d\bar{\mathbf{y}}}{d\tau} &=&  \left\langle\mathbf{f}\left(\bar{\mathbf{y}}+ \Phi(s) \mathbf{A}\bm{\psi}(\bar{\alpha}) \right)\right\rangle \label{slow_main1}, \\
\omega \frac{d \bar{\alpha}}{d\tau}  &=& \Omega, \label{slow_phase1}
\end{eqnarray}
\end{subequations}
where the angle brackets denote the averaging over the variable $s$. Note that in general the averaged Eqs. (\ref{slow1}) approximate solutions of the system (\ref{slow}) with accuracy $\mathbf{y}(\tau)=\bar{\mathbf{y}}(\tau)+O(\omega^{-1})$ on a time interval of the order $O(\omega)$ \cite{sand07}. However, here we are interested in stable periodic solutions of the system (\ref{slow1}). Then the above approximation is valid on the infinite time interval (cf.~\cite{burd07}, theorem 9.6.).

Further simplification can be made if we treat the components of the vector $\mathbf{A}$ as small parameters and
expand the function in the r.h.s. of Eq. (\ref{slow_main1})
in Taylor series
\begin{eqnarray}
& & \mathbf{f}\left(\bar{\mathbf{y}}+\Phi(s) \mathbf{A}\bm{\psi}(\bar{\alpha}) \right) = \mathbf{f}\left(\bar{\mathbf{y}} \right) +\Phi(s)\sum_{i=1}^n
\frac{\partial \mathbf{f}\left(\bar{\mathbf{y}} \right)}{\partial \bar{y}_i} A_i \psi_i(\bar{\alpha}) \nonumber \\
& & +\frac{\Phi^2(s)}{2}\sum_{i,j=1}^n \frac{\partial^2 \mathbf{f}\left(\bar{\mathbf{y}} \right)}{\partial \bar{y}_i \partial \bar{y}_j}A_i A_j \psi_i(\bar{\alpha}) \psi_j(\bar{\alpha}) + O\left(\mathbf{A}^3\right).
\label{taylor_ex}
\end{eqnarray}
Despite the fact that here we treat $A_i$ as small parameters, the product $\mathbf{K}=\omega \mathbf{A}$ can be large for large $\omega$ so that the perturbation in Eq. (\ref{main}) is not small. Using Eq. (\ref{taylor_ex}) we can perform explicitly the averaging in Eq. (\ref{slow_main1}). Then omitting the small term $O\left(\mathbf{A}^3\right)$ and returning to the original time scale, we get
\begin{eqnarray}
\dot{\bar{\mathbf{y}}}(t) &=& \mathbf{f}\left(\bar{\mathbf{y}}(t) \right) \nonumber \\
& & +\frac{\left\langle \Phi^2 \right\rangle}{2} \sum_{i,j=1}^n \frac{\partial^2 \mathbf{f}\left(\bar{\mathbf{y}}(t) \right)}{\partial \bar{y}_i \partial \bar{y}_j}A_i A_j \psi_i(\Omega t) \psi_j(\Omega t).
\label{main_aver}
\end{eqnarray}
Since the second term in the r.h.s. is  small [its order is $O\left(\mathbf{A}^2\right)$], we can treat this system by the conventional phase reduction method.  The unperturbed Eq.  (\ref{main_aver}) as well as the original Eq. (\ref{main}) has the stable limit cycle solution $\bar{\mathbf{y}}(t)=\bm{\xi}(t)$. The usual infinitesimal phase response curve (PRC) $\mathbf{z}(t)$ is defined as a $T_0$-periodic solution of the adjoint equation $\dot{\mathbf{z}}(t)=-[J(t)]^T\mathbf{z}(t)$, where $J(t)=D\mathbf{f}(\bm{\xi}(t))$ is the Jacobian of the free system evaluated on the limit cycle. As a result, we can write an equation  for the phase $\vartheta(t)$ of the system (\ref{main_aver}) as
\begin{equation}
\dot{\vartheta}(t) = 1+\frac{\left\langle \Phi^2 \right\rangle}{2}\mathbf{z}^T(\vartheta)\sum_{i,j=1}^n \frac{\partial^2 \mathbf{f}\left(\bm{\xi}(\vartheta) \right)}{\partial \xi_i \partial \xi_j}A_i A_j \psi_i(\Omega t) \psi_j(\Omega t).
\label{phas_dyn}
\end{equation}
In neuroscience, the coupling matrix has typically only one nonzero component, $\mathbf{K}=\textrm{diag}[K_1,0,\ldots,0]$. Then the Eq. (\ref{phas_dyn}) simplifies to
\begin{equation}
\dot{\vartheta}(t) = 1+\frac{\left\langle \Phi^2 \right\rangle}{2} A^2 z_{\mathrm{eff}}(\vartheta)\psi^2(\Omega t).
\label{phas_dyn_1}
\end{equation}
Here we skipped the subindexes in $A_1$ and $\psi_1$ and introduced an effective PRC as
\begin{equation}
z_{\mathrm{eff}}(\vartheta)=\mathbf{z}^T(\vartheta) \frac{\partial^2 \mathbf{f} \left(\bm{\xi}(\vartheta)\right)}{\partial \xi_1^2}.
\label{eff_prc}
\end{equation}
>From Eq.~(\ref{phas_dyn_1}) we can make two important conclusions: (i) the sign of the envelope $\psi$ does not affect the phase of the system and (ii) if $z_{\mathrm{eff}}(\vartheta)$ is positive (negative) on the whole interval $[0,T_0]$ then the entrainment of the oscillator is possible only for $\Omega>\Omega_0$ ($\Omega<\Omega_0$).

Below we present two specific examples to demonstrate the validity of our phase reduction theory.

\subsection{Example I: A Stuart-Landau oscillator}
We start from a simple example of a Stuart-Landau (SL) oscillator driven by the AMHF force:
\begin{subequations}
\label{sl}
\begin{eqnarray}
\dot{x}_1 &=& x_1\left[1-x_1^2-x_2^2 \right] -x_2+K \psi(\Omega t)\varphi(\omega t), \label{sl_1} \\
\dot{x}_2 &=& x_2\left[1-x_1^2-x_2^2 \right] +x_1. \label{sl_2}
\end{eqnarray}
\end{subequations}
Here the limit cycle and the conventional PRC of the free system can be found analytically: $\bm{\xi}(t)=[\cos(t),\sin(t)]^T$ and $\mathbf{z}(t)=[-\sin(t), \cos(t)]^T$. Then the effective PRC is $z_{\mathrm{eff}}(\vartheta)=2\sin(2\vartheta)$. We choose a particular waveform with the harmonic HF function $\varphi(\omega t)=\cos(\omega t)$ and the square wave envelope $\psi(\Omega t)=H(\sin(2\Omega t))$, where  $H(\cdot)$ is a Heaviside step function.

\begin{figure}[h!]
\centering\includegraphics[width=0.85\columnwidth]{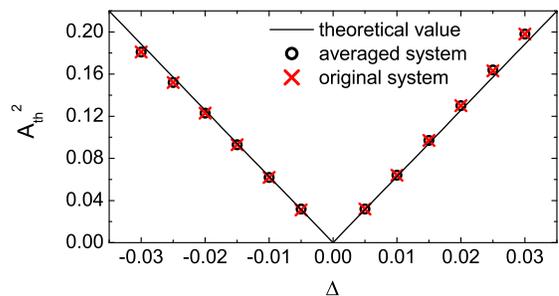}
\caption{\label{fig1} (Color online) The Arnold tongue of the SL system (\ref{sl}) for $\omega/ \Omega = 100$. Straight lines represent analytical Eq. (\ref{sl_amp}), black circles and red crosses show the numerical results derived from the averaged Eq.~(\ref{main_aver}) and  original Eq.~(\ref{sl}), respectively.}
\end{figure}

To derive an analytical expression for an entrainment threshold, we introduce a new phase variable $\chi(t)=\vartheta(t)-t\frac{\Omega}{\Omega_0}$ and rewrite the Eq. (\ref{phas_dyn_1}) in the form
\begin{equation}
\dot{\chi} = -\Delta+\frac{\left\langle \Phi^2 \right\rangle}{2} A^2 z_{\mathrm{eff}}\left(\chi+t\frac{\Omega}{\Omega_0}\right)\psi^2(\Omega t),
\label{phase}
\end{equation}
where
\begin{equation}
\Delta=\Omega/\Omega_0-1
\label{mismatch}
\end{equation}
is the frequency mismatch. The r.h.s. of Eq. (\ref{phase}) is a $T$-periodic function, where $T=2\pi/\Omega$ is the envelope period. Assuming that the frequency mismatch $\Delta$ is a small parameter of the same order $O(A^2)$ as the second term in the Eq.~(\ref{phase}), we can treat this system by the method of averaging. Denoting the variable of the averaged system as $\bar{\chi}$, we get an equation
\begin{equation}
\dot{\bar{\chi}} = -\Delta+\frac{\left\langle \Phi^2 \right\rangle}{2} A^2 G( \bar{\chi}),
\label{phase_avr}
\end{equation}
where $G( \bar{\chi})$ is a $T_0$-periodic function defined as:
\begin{eqnarray}
G(\bar{\chi}) &=& \frac{1}{T}\int_0^{T}z_{\mathrm{eff}}\left(\bar{\chi}+s\frac{\Omega}{\Omega_0}\right)\psi^2(\Omega s)ds \nonumber\\
 &=& \frac{1}{T_0}\int_0^{T_0}z_{\mathrm{eff}}\left(\bar{\chi}+s\right)\psi^2(\Omega_0 s)ds.
\label{for_g}
\end{eqnarray}
The Eq. (\ref{phase_avr}) approximates the solution of Eq. (\ref{phase}) with the accuracy $O(A^2)$, $\bar{\chi}(t)=\chi(t)+O(A^2)$. The entrainment of the oscillator to the envelope frequency $\Omega$ takes place when the system (\ref{phase_avr}) possesses a stable fixed point. The maximal and minimal values of the function $G(\bar{\chi})$ define the threshold amplitude $A=A_{\textrm{th}}$ at which the entrainment appearers.  For the given waveform, we have $\left\langle \Phi^2 \right\rangle = 1/2$, $z_{\mathrm{eff}}(\vartheta)=2\sin(2\vartheta)$ and $\psi(t)=H(\sin(2 t))$, so that the maximal and minimal values of the function $G(\bar{\chi})$ are: $\max[G(\bar{\chi})]=G(0)=2/\pi$ and $\min[G(\bar{\chi})]=G(\pi/2)=-2/\pi$. Inserting these values into Eq.~(\ref{phase_avr}) and equating the r.h.s to zero, we get the threshold amplitude
\begin{equation}
A_{\mathrm{th}}  = \sqrt{2\pi |\Delta|}.
\label{sl_amp}
\end{equation}
As is seen from FIG.~\ref{fig1}, the Arnold tongue computed numerically from the averaged Eq. (\ref{main_aver}) and  original Eq. (\ref{sl}) is in good agreement with the analytical result (\ref{sl_amp}).

\subsection{Example II: A Morris-Lecar model neuron}

Now we apply our phase reduction theory to a Morris-Lecar \cite{morris81}  model neuron subjected to the AMHF force:
\begin{subequations}
\label{ml}
\begin{eqnarray}
C\dot{V} &=& -g_{Ca} m_{\infty}(V)(V-V_{Ca})-g_K w(V-V_K)\nonumber\\
&-& g_l(V-V_l)+I+K \psi(\Omega t)\varphi(\omega t), \label{ml_1} \\
\dot{w} &=& \phi [w_{\infty}(V)-w]/\tau_w(V), \label{ml_2}
\end{eqnarray}
\end{subequations}
where $m_{\infty}(V)=0.5 \left\lbrace 1+\tanh[(V-V_1)/V_2] \right\rbrace$, $w_{\infty}(V)=0.5 \left\lbrace 1+\tanh[(V-V_3)/V_4] \right\rbrace$ and $\tau_w (V)=1/\cosh[(V-V_3)/(2V_4)]$. The parameter values are: $C=5.0$ $\mu$F/cm$^2$, $g_{Ca}=4.0$ $\mu$S/cm$^2$, $g_{K}=8.0$ $\mu$S/cm$^2$, $g_l=2.0$ $\mu$S/cm$^2$, $V_{Ca}=120$ mV, $V_{K}=-80$ mV, $V_{l}=-60$ mV, $V_{1}=-1.2$ mV, $V_{2}=18.0$ mV, $V_{3}=12$ mV, $V_{4}=17.4$ mV, $\phi=1/15$ ms$^{-1}$ and $I=40.0$ $\mu$A/cm$^2$.

\begin{figure}[h!]
\centering\includegraphics[width=0.85\columnwidth]{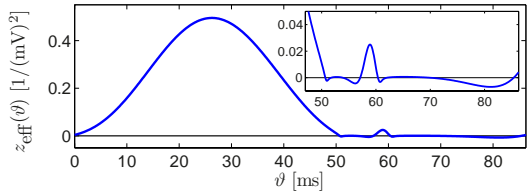}
\caption{\label{fig1_supp} The effective phase response curve for the Morris-Lecar neuron model~(\ref{ml}).  The inset shows an enlarged segment of the effective PRC, where it has negative values.}
\end{figure}

For the given values of the parameters, the free neuron fires with the period $T_0\approx 86.27$ ms. The numerically computed effective PRC is depicted in FIG.~\ref{fig1_supp}. We see that it is positive almost on the whole interval and there are some regions of $\vartheta$ where this function has very small negative values.
This means that the entrainment of the neuron is effective only for the positive frequency mismatch $\Delta>0$. We  choose  the HF function in the form of harmonic signal $\varphi=\cos(\omega t)$ with $\omega=100\Omega$ and verify our theory for two different waveforms of the envelope: (i) the harmonic wave envelope $\psi(\Omega t)= (1-\cos(\Omega t))/2$ and (ii) the square wave envelope $\psi(\Omega t))= H(\sin(\Omega t))$, which a half of the period is equal to 1 and another half is equal to 0.
\begin{figure}[h!]
\centering\includegraphics[width=0.85\columnwidth]{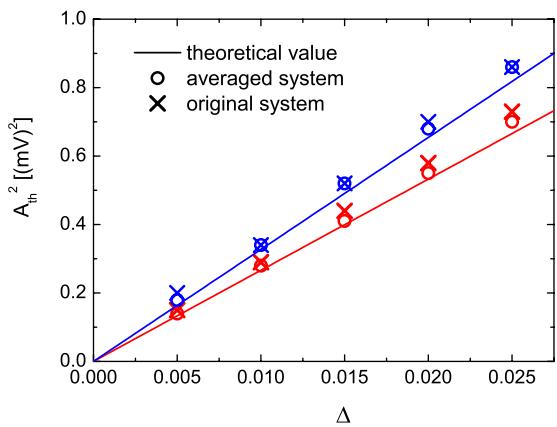}
\caption{\label{fig2_supp} (Color online) The Arnold tongues for the Morris-Lecar neuron~(\ref{ml}). The blue color represents the harmonic wave envelope $\psi(\Omega t)= (1-\cos(\Omega t))/2$, while the red color corresponds to the square wave envelope $\psi(\Omega t)= H(\sin(\Omega t))$. The strait lines show the theoretical values defined by Eqs.~(\ref{a_th_ml_harm}) and (\ref{a_th_ml_square}), circles show the numerical results obtained from averaged system (\ref{main_aver})  and the crosses represent the results of direct numerical simulation of the original system~(\ref{ml}).}
\end{figure}

For the given envelopes, we numerically estimated the function $G( \bar{\chi})$ defined by Eq. (\ref{for_g}) and found that it is everywhere positive. Therefore, the entrainment is impossible for $\Delta<0$. The theoretical value  of the threshold amplitude  can be derived from Eq. (\ref{phase_avr}) by replacing $G(\bar{\chi})$ with the maximal value $\max[G(\bar{\chi})]$ and equating the right hand side to zero. For the harmonic wave envelope we get:
\begin{equation}
A_{\textrm{th}}^2 = \left\lbrace
\begin{array}{l}
32.72 \Delta \; \textrm{when} \; \Delta>0 \\
\infty  \; \textrm{when} \; \Delta<0
\end{array} \right. .
\label{a_th_ml_harm}
\end{equation}
Similarly, the threshold amplitude for the square wave envelope is given by
\begin{equation}
A_{\textrm{th}}^2 = \left\lbrace
\begin{array}{l}
26.64 \Delta \; \textrm{when} \; \Delta>0 \\
\infty  \; \textrm{when} \; \Delta<0
\end{array} \right. .
\label{a_th_ml_square}
\end{equation}
In FIG. \ref{fig2_supp}, these theoretical values are compared with the results of numerical simulation of the averaged Eq. (\ref{main_aver})  and the original system (\ref{ml}). For  both waveforms, our phase reduction theory predicts correctly the results of direct numerical simulations of the original system.

\begin{figure}[h!]
\centering\includegraphics[width=0.85\columnwidth]{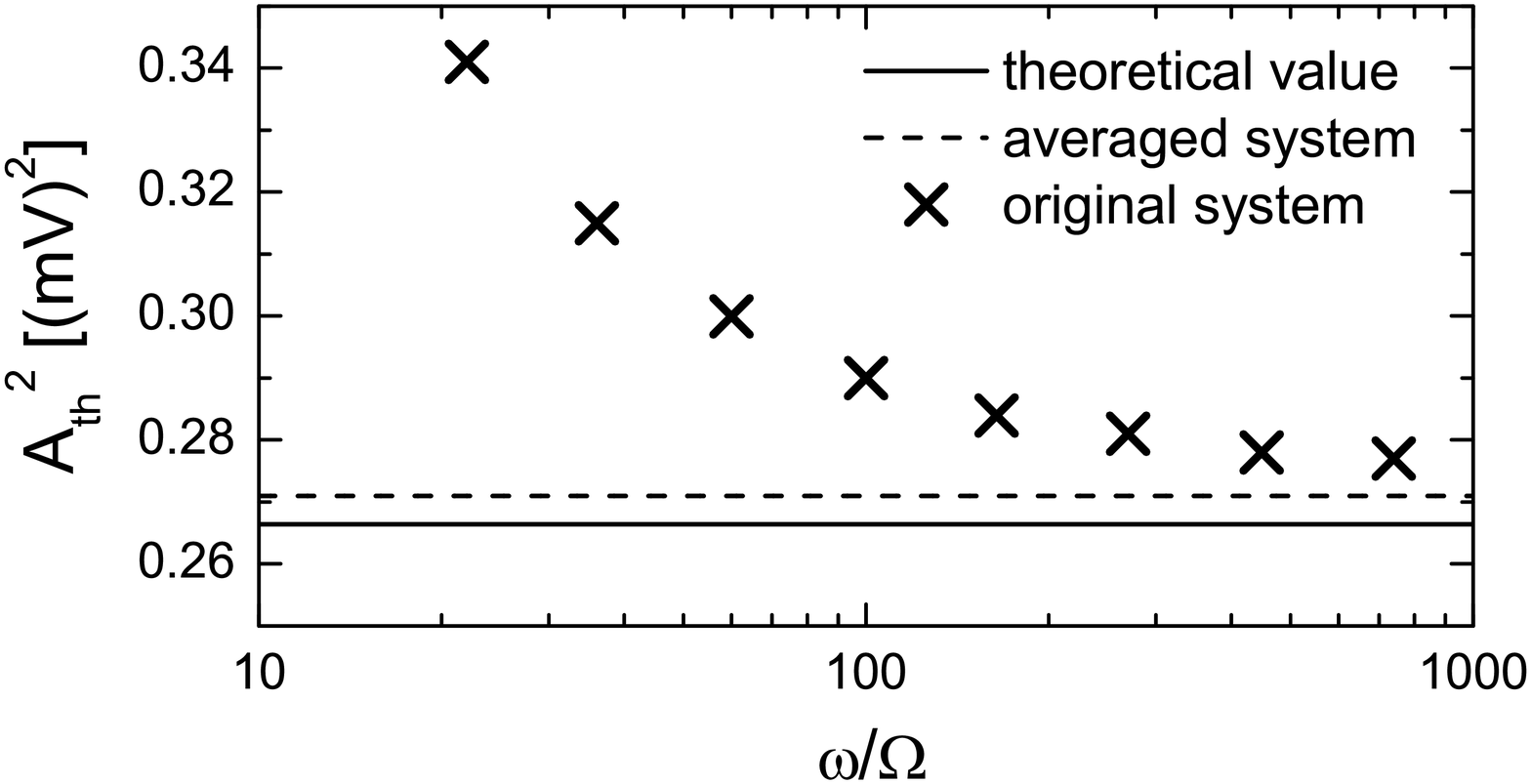}
\caption{\label{fig4_delta_fix} The threshold amplitude as the function of the carrier frequency for the Morris-Lecar neuron~(\ref{ml}). The numerical computations are performed for the fixed frequency mismatch $\Delta=0.01$ using the square wave envelope $\psi(\Omega t)= H(\sin(\Omega t))$ with the varying carrier frequency $\omega$.  The solid line shows the theoretical value obtained from the Eq.~(\ref{a_th_ml_square}), while the dashed line is computed from the averaged system~(\ref{main_aver}). The crosses represent the results of direct numerical simulation of the original system~(\ref{ml}).}
\end{figure}

In order to demonstrate how the solution of the averaged system (\ref{main_aver}) approaches the solution of the original system (\ref{ml}) with the increase of $\omega$, we fixed the frequency mismatch $\Delta=0.01$ and computed the threshold amplitude $A_{\textrm{th}}$. The results for the square wave envelope with the varying carrier frequency $\omega$ are presented in FIG.~\ref{fig4_delta_fix}. We see that the results obtained from the original system (\ref{ml}) converge to the value derived  from the averaged system (\ref{main_aver}), while the latter  approaches the theoretical value (\ref{a_th_ml_square}) in the limit $\Delta \rightarrow 0$.

\section{\label{sec3} The AMHF waveform optimization}

The phase Eq.~(\ref{phas_dyn_1}) is helpful to solve the waveform optimization problem. For the fixed frequencies $\omega$ and $\Omega$, we are seeking to find the optimal waveforms $\varphi(\omega t)$ and $\psi(\Omega t)$, which provide an entrainment of a given oscillator to the envelope frequency $\Omega$ with minimal power. We assume that the external force is restricted by some value $I_0$, so that $\left| K \psi(\Omega t) \varphi(\omega t) \right| \leq I_0$ holds for any time. It means that the amplitude $A$ cannot exceed the value $I_0/\omega$. To solve this problem, we invoke the Pontryagin maximum principle \cite{pontryagin1962}. Here we present only the main results, while the details are provided in the Appendix.

Assuming that the envelope $\psi(\Omega t)$ is a slowly varying function on the HF period $2\pi/\omega$, the power $P=(\Omega/2\pi)\int_0^{2\pi/\Omega}\left[K \psi(\Omega t) \varphi(\omega t)\right]^2 dt$ of the perturbation can be approximated as a product of two factors:
\begin{equation}
P = \left( \frac{ \omega^2}{2\pi} \int_0^{2\pi} A^2\psi^2 (s)ds\right) \left( \frac{1}{2\pi} \int_0^{2\pi} \varphi^2 (s)ds\right).
\label{power}
\end{equation}
We denote the first and the second factor  as $P_{\Omega}$ and $P_{\omega}$, respectively. Since $P_{\Omega}$ depends only on $A\psi$ and $P_{\omega}$ depends only on  $\varphi$, the problems of the $A\psi$ and $\varphi$ waveforms optimization can be analyzed separately. We show (see the Appendix) that the optimal HF waveform (which we mark by an asterisk) is the harmonic function $\varphi^*(s)=\sin(s+\beta)$ and thus $P_{\omega}=1/2$. If the harmonic wave is replaced by the square wave $\varphi(s)=\text{sgn}(\sin(s+\beta))$ then the threshold power necessary to achieve an entrainment will increase by the factor $1.22$.

The optimal waveform of the envelope represents a switching function with two possible values $\psi^*=1$ (switched on) and $\psi^*=0$ (switched off).  The time intervals where the perturbation is switched on and off are defined with the help of two auxiliary functions
\begin{subequations}
\label{mpm}
\begin{eqnarray}
M^{+}(u) &=& \left\langle H(z_\textrm{eff}(\vartheta)-u) z_\textrm{eff}(\vartheta) \right\rangle \: \textrm{when} \: \Delta > 0, \label{m_plus} \\
M^{-}(u) &=& \left\langle H(u-z_\textrm{eff}(\vartheta)) z_\textrm{eff}(\vartheta) \right\rangle \: \textrm{when} \: \Delta < 0, \label{m_minus}
\end{eqnarray}
\end{subequations}
where the angle brackets denote the averaging over $\vartheta$. The both functions $M^{\pm}(u)$ are monotonically decreasing functions. The function $M^{+}(u)$ ($M^{-}(u)$) is determined only for the positive (negative) $u$ and turns to zero at the point $u^+_c=\max [z_\textrm{eff}(\vartheta)] $ ($u^-_c=\min [z_\textrm{eff}(\vartheta)] $) [cf. FIG.~\ref{fig5}(c)]. Using these functions, we determine a point $u_0$ where
\begin{equation}
M^{\pm}( u_0) = \frac{2 \omega^2 \Delta}{\left\langle \Phi^2 \right\rangle I_0^2}
\label{for_e}
\end{equation}
and then define the optimal envelope as:
\begin{equation}
\psi^*(\Omega_0 \vartheta) = \left\lbrace
\begin{array}{l}
H(\Delta) \; \textrm{when} \; z_\textrm{eff}(\vartheta)>u_0 \\
H(-\Delta) \; \textrm{when} \; z_\textrm{eff}(\vartheta)<u_0
\end{array} \right. .
\label{opt_psi}
\end{equation}
The optimal value of the amplitude $A$ is its maximal allowable value $A^*=I_0/\omega$. Note that the entrainment is possible only when  $I_0>I_{\textrm{cr}}=\omega\left[2  \Delta/\left\langle \Phi^2 \right\rangle M^{\pm}(0)\right]^{1/2}$.
The waveform $A^*\psi^*(\Omega_0 \vartheta)$ provides an entrainment of the oscillator to the envelope frequency $\Omega$ with the lowest possible power $P_{\Omega}=I_0^2 N^{\pm}( u_0)$, where the functions $N^{\pm}(u)$ are
\begin{subequations}
\label{npm}
\begin{eqnarray}
N^{+}(u) &=& \left\langle H(z_\textrm{eff}(\vartheta)-u) \right\rangle \: \textrm{when} \: \Delta > 0, \label{n_plus} \\
N^{-}(u) &=& \left\langle H(u-z_\textrm{eff}(\vartheta)) \right\rangle \: \textrm{when} \: \Delta < 0. \label{n_minus}
\end{eqnarray}
\end{subequations}

For large $I_0$, the optimality of the waveform (\ref{opt_psi}) has a clear qualitative explanation. Assume that the frequency mismatch is positive, $\Delta>0$. Then for $I_0\to \infty$, the point $u_0$ approaches the maximum $u_c^+$ of the curve $z_\textrm{eff}(\vartheta)$ and the waveform $A^*\psi^*(\Omega_0\vartheta)$ turns into a narrow high pulse located at the point $\vartheta$ where this maximum is reached, i.e., the whole power of the perturbation is consumed at this point. >From Eq. (\ref{phas_dyn_1}) it follows that such a waveform provides the maximal increase of the oscillator phase during the period of oscillations.

\subsection*{Example: A FitzHugh-Nagumo model neuron}
We demonstrate the waveform optimization theory with the specific example of a  FitzHugh-Nagumo (FHN) \cite{Fitzhugh61, *Nag62} neuron  driven by the AMHF force:
\begin{subequations}
\label{fhn}
\begin{eqnarray}
\dot{x}_1 &=& x_1-x_1^3/3-x_2+a+K \psi(\Omega t)\varphi(\omega t), \label{fhn_1} \\
\dot{x}_2 &=& \varepsilon \left( x_1+b_0-b_1 x_2 \right). \label{fhn_2}
\end{eqnarray}
\end{subequations}

\begin{figure}[t]
\centering\includegraphics[width=0.95\columnwidth]{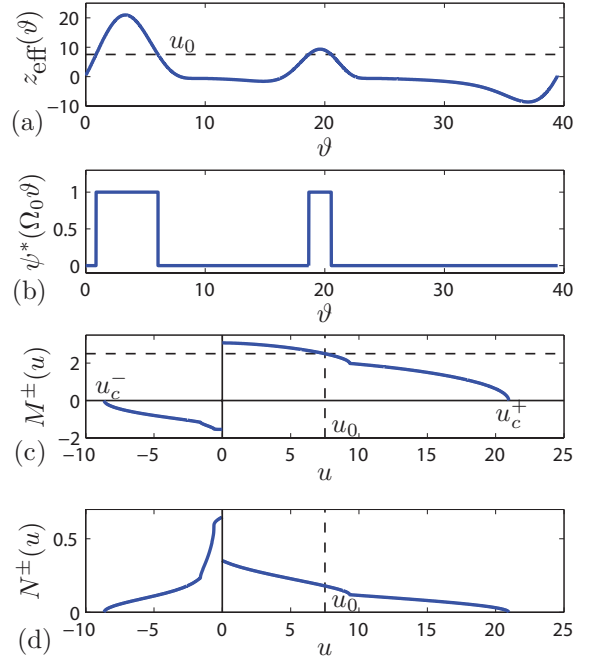}
\caption{\label{fig5} Waveform optimization for the FHN neuron (\ref{fhn}): (a) -- the effective PRC, (b) -- an example of optimal envelope for $\Delta=0.1$, (c) -- the functions $M^{\pm}(u)$ defined by Eqs. (\ref{mpm}) and (d) -- the functions $N^{\pm}(u)$ defined by Eqs. (\ref{npm}).}
\end{figure}

For the fixed values of the parameters $a=0.5$, $\varepsilon=0.08$, $b_0=0.7$ and $b_1=0.8$, the free neuron fires with the period  $T_0 \approx 39.47$. Numerically computed effective PRC is depicted in FIG.~\ref{fig5}(a). We take an optimal HF function $\varphi^*(\omega t)$ in the form of harmonic signal with the frequency $\omega = 1000\Omega$ and choose $I_0=70$. An example of optimal envelope for the fixed  $\Delta=0.1$ is shown in panel (b). We present a graphical illustration of how the envelope is constructed. For the given values of parameters, the r.h.s. of Eq.~(\ref{for_e}) is equal to $2.5$. This value  is depicted as a horizontal dashed line in panel (c). Its intersection with the curve $M^+(u)$ gives the value $u_0$, which is represented by a vertical dashed line. Then we depict the value $u_0$ as a horizontal dashed line in panel (a). Finally, the optimal envelope $\psi^*(\Omega_0\vartheta)$ is equal to 1 in the regions of $\vartheta$ where $z_{\textrm{eff}}(\vartheta)>u_0$ and is equal to 0 otherwise.

\begin{figure}[h]
\centering\includegraphics[width=0.85\columnwidth]{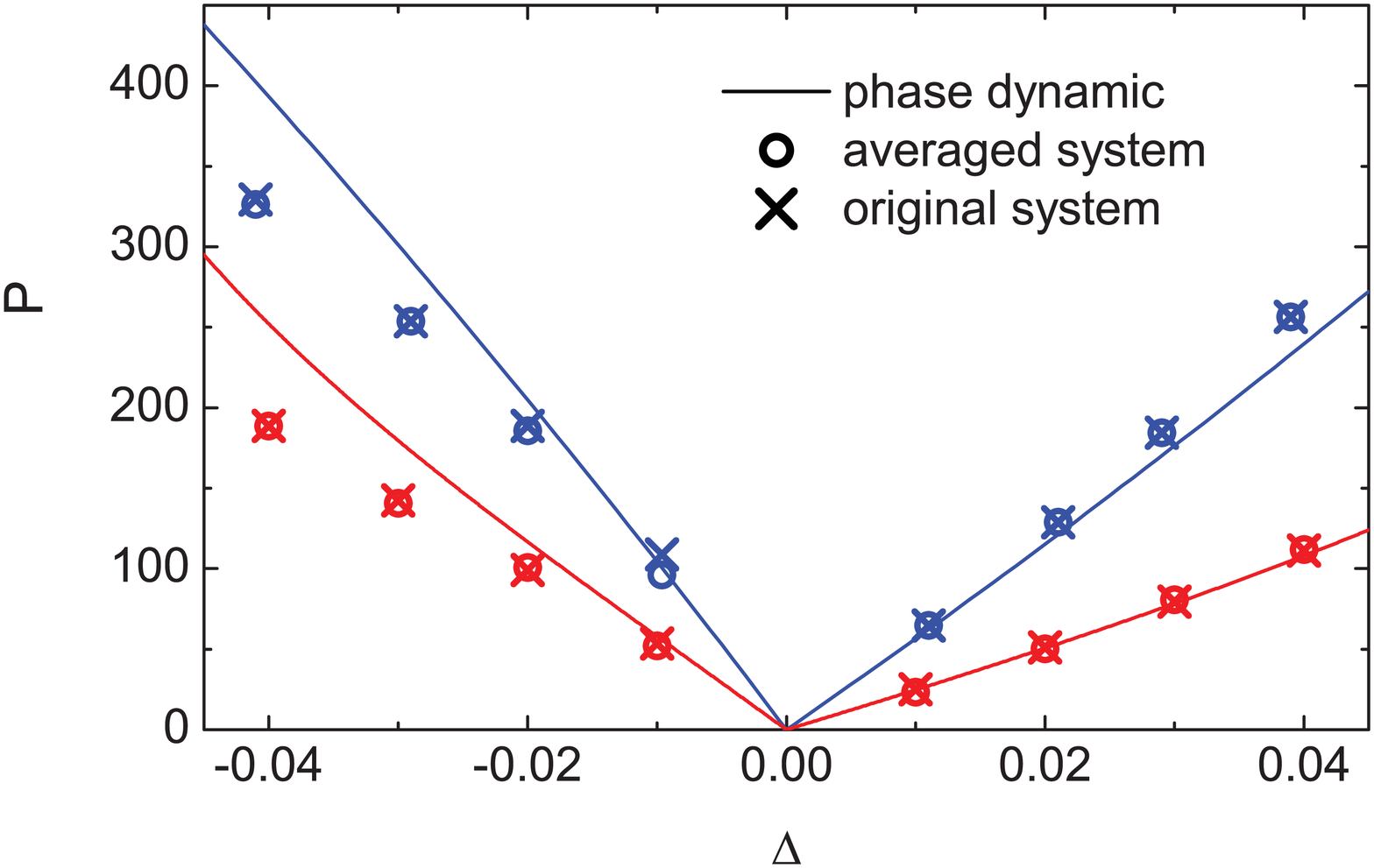}
\caption{\label{fig6} (Color online) The Arnold tongues of the FHN system~(\ref{fhn}). The red and blue colors show the results obtained with the optimal  $\psi^*$ and ``quarter'' $\psi_{1/4}$ envelope, respectively. Solid curves are derived from the phase Eq. (\ref{phas_dyn_1}), circles represent the results of the averaged Eq. (\ref{main_aver}) and the crosses show the results obtained from the original system (\ref{fhn}). When computing the solid curves for the optimal envelope, we fixed $I_0=70$, while for circles and crosses, at each given $\Delta$, we used the same waveform as for the solid curve and varied slightly $I_0$ until the entrainment threshold was reached.}
\end{figure}

In FIG. \ref{fig6}, we compare the Arnold tongues of the FHN model obtained with two different envelopes: (i) the optimal  envelope $\psi^*$ defined by Eq.~(\ref{opt_psi}) and (ii) a non-optimal, ``quarter'' envelope $\psi_{1/4}$, which a quarter of the period is equal to 1 and the rest part is equal to 0. In both cases we take the HF carrier signal $\varphi(\omega t)$ as a harmonic function. The minimal power necessary to attain an entrainment of the oscillator has been estimated by three different methods, namely, using the phase Eq. (\ref{phas_dyn_1}), the averaged Eq. (\ref{main_aver}) and the original system (\ref{fhn}). The simulations confirm the advantage of the optimal envelope, since it provides the entrainment with less power as compared to the ``quarter'' envelope.

\section{\label{sec4} Conclusions}
In conclusion, we have developed the phase reduction theory for a limit cycle oscillator driven by a strong amplitude-modulated high-frequency force and found an optimal waveform that ensures an entrainment of the oscillator with  minimal power. Our findings are relevant to design of mild neurostimulation protocols for treatment of neurological diseases.

\begin{acknowledgments}
This research was funded by the European Social Fund under
the Global Grant measure (grant No.~VP1-3.1-\v{S}MM-07-K-01-025).
\end{acknowledgments}

\appendix*
\section{The AMHF waveform optimization}
According to Eq. (\ref{power}), the power of the AMHF perturbation can be presented as a product of two factors $P=P_\Omega P_\omega$, where
\begin{equation}
P_\Omega  = \left( \frac{\omega^2}{2\pi} \int_0^{2\pi} A^2\psi^2(s)ds\right), \, P_\omega=\left( \frac{1}{2\pi} \int_0^{2\pi} \varphi^2(s)ds\right).
\label{pow}
\end{equation}
For the fixed frequencies $\omega$ and $\Omega$, we are seeking to find the optimal waveforms $\varphi(\omega t)$ and $\psi(\Omega t)$ as well as the optimal value of the amplitude $A$, which provide an entrainment of a given oscillator to the envelope frequency $\Omega$ with minimal power $P$. The dynamics of the oscillator is defined by Eq. (\ref{phas_dyn_1}) given in the main text. For clarity of the presentation,  here we rewrite this equation:
\begin{equation}
\dot{\vartheta}(t) = 1+\frac{\left\langle \Phi^2 \right\rangle}{2} A^2 \psi^2(\Omega t)z_{\mathrm{eff}}(\vartheta).
\label{phas_dyn_1a}
\end{equation}
The entrainment takes place if the system (\ref{phas_dyn_1a}) admits a solution with the boundary conditions
\begin{subequations}
\label{bound}
\begin{eqnarray}
\vartheta(0) &=& 0, \label{bound_1} \\
\vartheta(T) &=& T_0, \label{bound_2}
\end{eqnarray}
\end{subequations}
where $T_0=2\pi/\Omega_0$ is the natural period of the oscillator and $T=2\pi/\Omega$ is the period of the envelope. The main conditions of the optimization are  as follows. The both functions $\varphi(s)$ and $\psi(s)$ are $2\pi$-periodic with respect to the variable $s$ and their values lie in the interval $[-1, 1]$. The function $\varphi(s)$ satisfies the charge-balanced condition $\int_0^{2\pi}\varphi(s)ds=0$. The external force is restricted by some value $I_0$, so that $\left| K \psi(\Omega t) \varphi(\omega t) \right| \leq I_0$ holds for any time. Thus the the amplitude $A$ is restricted by the interval $A\in [0, I_0/\omega]$.

Note that in all equations, the function $\psi$ and the amplitude $A$ appear as a product $A \psi$ and thus the variation of $A$ and $\psi$ can be considered as a variation of a new function $\Psi(s)=A\psi(s)$. The function $\Psi(s)$ admits the variation of both the amplitude and the waveform. This is in contrast to the function $\psi(s)$, which has a fixed amplitude and admits the variation  of only the waveform.  Let's say, we have found such  $\varphi$ and $A\psi$ that satisfy Eq. (\ref{phas_dyn_1a}) with the boundary conditions (\ref{bound}). For the given $\varphi$, let us denote the value of $\left\langle \Phi^2 \right\rangle$ by $\left\langle \Phi^2 \right\rangle \equiv B$. First we fix $A\psi$ and $\left\langle \Phi^2 \right\rangle$ and vary $\varphi$ in order to minimize the power. Since the power functional is the product of two functionals $P=P_\Omega[A\psi]P_\omega[\varphi]$, our first problem is to minimize $P_\omega[\varphi]$ for the fixed $\left\langle \Phi^2 \right\rangle$. This allows us to find an optimal high frequency waveform $\varphi^*$. In the second stage, we fix $\varphi=\varphi^*$ and vary  $A\psi$  in order to minimize the functional $P_\Omega[A\psi]$.

The next two sections are devoted to the solution of these two separate problems.

\subsection*{High frequency waveform optimization}

We start from optimization of the high frequency waveform $\varphi$.  For a given value $\left\langle \Phi^2 \right\rangle = B$, we are seeking to minimize the functional $P_\omega[\varphi]$ with the constrains $\int_0^{2\pi}\varphi(s)ds=0$ and $\varphi(s+2\pi)=\varphi(s)$.  We also require that the maximum of the function $\varphi(s)$ is equal to $1$ and the minimum is not bellow than $-1$ (see the main text). Using Eq. (\ref{antider}), the term $\left\langle \Phi^2 \right\rangle$ can be written as
\begin{equation}
\left\langle \Phi^2 \right\rangle =  \left\langle \Phi^2_1 \right\rangle-\left\langle \Phi_1 \right\rangle^2 .
\label{fi_sq}
\end{equation}
We rewrite the function $\Phi_1(s)=\int_0^s\varphi(s')ds'$ in the form $\Phi_1(s)=\int_0^{2\pi}[1-H(s'-s)]\varphi(s')ds'$, where $H(\cdot)$ is  the Heaviside step function. Now we can write down the functional
\begin{widetext}
\begin{eqnarray}
& & J[\varphi] = P_\omega[\varphi]+\lambda_1 \int\limits_0^{2\pi}\varphi(s)ds+\lambda_2 \left\lbrace \left\langle \Phi^2 \right\rangle-B\right\rbrace=\frac{1}{2\pi}\int\limits_0^{2\pi}\varphi^2(s)ds+\lambda_1\int\limits_0^{2\pi}\varphi(s)ds \nonumber \\
& & +\lambda_2 \left\lbrace \frac{1}{2\pi}\int\limits_0^{2\pi} \left( \int\limits_0^{2\pi}[1-H(s-t)]\varphi(s)ds\right)^2 dt-\left( \frac{1}{2\pi}\int\limits_0^{2\pi}\int\limits_0^{2\pi}[1-H(s-t)]\varphi(s)ds dt\right)^2 -B\right\rbrace, \nonumber
\label{aux_func}
\end{eqnarray}
\end{widetext}
which we aim to minimize. Here $\lambda_1$ and $\lambda_2$ are the Lagrange multipliers.  Equating the first variation of the functional to zero, we obtain:
\begin{eqnarray}
& & \frac{2\varphi(s)}{2\pi}+\lambda_1 \nonumber \\
& & +\frac{\lambda_2}{2\pi}\int\limits_0\limits^{2\pi}\int\limits_0\limits^{2\pi} 2[1-H(y-t)][1-H(s-t)]\varphi(y)dydt -\frac{\lambda_2}{(2\pi)^{2}} \nonumber \\
& & \times\int\limits_0^{2\pi}\int\limits_0^{2\pi}\int\limits_0^{2\pi} 2[1-H(y-z)][1-H(s-t)]\varphi(y)dydzdt = 0. \nonumber
\label{variat}
\end{eqnarray}
This is a rather complicated integral equation. However, by differentiating  this equation two times with respect to the variable $s$, we come to the differential equation:
\begin{equation}
\varphi''(s) -\lambda_2 \varphi(s)=0 .
\label{fi_diff}
\end{equation}
Since the function $\varphi(s)$ is $2\pi$-periodic and its maximum is equal to 1, we obtain that $\lambda_2=-1$ and  $\varphi(s)=\sin(s+\beta)$. Thus the optimal HF waveform (which we mark by an asterisk) is the harmonic signal $\varphi^*(s)=\sin(s+\beta)$. Note that this function automatically satisfies the charge-balanced condition $\int_0^{2\pi}\varphi^*(s)ds=0$. Also, it follows that  $B=1/2$. We have obtained the defined value of $B$ due to the fixed amplitude of the function $\varphi$.  Finally, the minimal value of the functional $P_\omega[\varphi]$ is:
\begin{equation}
P_\omega[\varphi^*]=1/2 .
\label{power_hf}
\end{equation}

\subsection*{Optimization of the envelope waveform}
Now we consider the problem of optimization of the waveform $A\psi$. Our aim is  to attain an entrainment of the perturbed oscillator to the envelope frequency $\Omega$ with the minimal value of the functional $P_\Omega[A\psi]$. We recall that the envelope $\psi(s)$ is a $2\pi$-periodic function whose values are in the interval $-1\le \psi(s)\le 1$ and the maximum of $\psi^2(s)$ is equal to $1$. Also, the external perturbation never exceeds some predefined value $I_0$, i.e., $|K\psi(\Omega t)\varphi(\omega t)|\leq I_0$ or $|\omega A\psi(\Omega t)|\leq I_0$ for any time. From here it follows that $A\in[0,I_0/\omega]$.

To minimize the envelope's power functional
\begin{equation}
P_\Omega[A\psi]=\frac{ \omega^2}{T}\int\limits_0^{T}A^2 \psi^2(\Omega t)dt,
\label{power_lf}
\end{equation}
with the above listed conditions, we refer to Pontriagin's theory \cite{pontryagin1962}. To this end we introduce the Lagrangian as $\mathcal{L}(A\psi)=A^2\psi^2(\Omega t)\omega^2/T$ and define the Hamiltonian of the system as $\mathcal{H}\left( \vartheta, A\psi, p \right)=p\dot{\vartheta}-\mathcal{L}(A\psi)$ or
\begin{eqnarray}
& & \mathcal{H}\left( \vartheta(t), A\psi(\Omega t), p(t) \right) = p(t) \nonumber \\
& & +A^2 \psi^2(\Omega t)\left[\frac{\left\langle \Phi^2 \right\rangle}{2} z_{\mathrm{eff}}(\vartheta) p(t) -\frac{\omega^2}{T}\right].
\label{hamil}
\end{eqnarray}
We denote the optimal trajectory (where $P_\Omega[A\psi]$ is minimal) with an asterisk: $\vartheta^*(t)$, $A^*\psi^*(\Omega t)$ and $p^*(t)$.
The Pontryagin maximum principle states that the Hamiltonian is constant on the optimal trajectory and this constant is the maximum possible value of the Hamiltonian.  Applying this principle to Eq. (\ref{hamil}), we easily derive the optimal waveform of the envelope
\begin{equation}
\psi^*(\Omega t) = \left\lbrace
\begin{array}{l}
1 \; \textrm{when} \; z_{\mathrm{eff}}(\vartheta^*)p^*(t)>\frac{2 \omega^2}{\left\langle \Phi^2 \right\rangle T} \\
0  \; \textrm{when} \; z_{\mathrm{eff}}(\vartheta^*)p^*(t)<\frac{2 \omega^2}{\left\langle \Phi^2 \right\rangle T}
\end{array} \right.
\label{opt_env}
\end{equation}
and obtain that the optimal value of the amplitude is its maximal allowable value, $A^*=I_0/\omega$. Let us denote the maximum constant value of the Hamiltonian as $\frac{2 \omega^2}{\left\langle \Phi^2 \right\rangle T u_0}$, i.e., $\mathcal{H}(\vartheta^*,A^*\psi^*,p^*)=\frac{2 \omega^2 }{\left\langle \Phi^2 \right\rangle T u_0}$. Here $u_0$ is some constant, whose value will be determined later. Then in time intervals, where $\psi^*(\Omega t)$ is equal to zero, we have $p^*(t)=\frac{2 \omega^2 }{\left\langle \Phi^2 \right\rangle T u_0}$. Therefore the second condition of the Eq.~(\ref{opt_env}) simplifies to $z_{\mathrm{eff}}(\vartheta^*)/u_0 <1$. The first condition of the Eq.~(\ref{opt_env})  can be simplified as well.  We substitute $\psi^*(\Omega t)=1$ and $A^*=I_0/\omega$ into the Eq.~(\ref{hamil}) and find $p^*(t)$. Then  inserting the obtained $p^*(t)$ into the first condition, we find that it transforms to $z_{\mathrm{eff}}(\vartheta^*)/u_0>1$. Finally, the Eq.~(\ref{opt_env}) simplifies to:
\begin{equation}
\psi^*(\Omega t) = \left\lbrace
\begin{array}{l}
1 \; \textrm{when} \; z_{\mathrm{eff}}(\vartheta^*)/u_0>1 \\
0  \; \textrm{when} \; z_{\mathrm{eff}}(\vartheta^*)/u_0<1
\end{array} \right. .
\label{opt_env1}
\end{equation}

Now using the Eq. (\ref{phas_dyn_1a}) and conditions~(\ref{bound}), we can define the constant $u_0$. For the positive frequency mismatch $\Delta>0$, we need to increase the phase velocity $\dot\vartheta$ in order to attain an entrainment. Therefore, we have to switch on the perturbation, $\psi^*(\Omega t)=1$, in the time intervals  where $z_{\mathrm{eff}}(\vartheta^*(t))$ is positive [see Eq. (\ref{phas_dyn_1a})]. For $\Delta<0$, the phase velocity has to decrease, and thus the perturbation has to be switched on, $\psi^*(\Omega t)=1$, in the time intervals  where $z_{\mathrm{eff}}(\vartheta^*(t))$ is negative. This means that the the constant $u_0$ has to be of the same sign as the mismatch $\Delta$. From Eq. (\ref{phas_dyn_1a}) and conditions~(\ref{bound}), we obtain
\begin{equation}
\int\limits_0^{T_0}\frac{d\vartheta^*}{1+\frac{\left\langle \Phi^2 \right\rangle I_0^2}{2\omega^2}z_{\mathrm{eff}}(\vartheta^*)\psi^{*2}(\Omega t)}= \int\limits_0^T dt.
\label{phase_dyn}
\end{equation}
Taking into account that $\omega^{-2}$ is a small parameter, we   expand the l.h.s. of the Eq.~(\ref{phase_dyn}) in Taylor series. Then  discarding the terms $O(\omega^{-4})$, we get:
\begin{equation}
T_0 + \frac{\left\langle \Phi^2 \right\rangle I_0^2}{2\omega^2} \int\limits_0^{T_0}z_{\mathrm{eff}}(\vartheta^*)\psi^{*2}(\Omega t) d\vartheta^*= T.
\label{phase_dyn1}
\end{equation}
By introducing the auxiliary functions (\ref{mpm})
the Eq.~(\ref{phase_dyn1}) can be rewritten as:
\begin{equation}
\frac{\left\langle \Phi^2 \right\rangle I_0^2}{2\omega^2} M^{\pm}(u_0)= \frac{T-T_0}{T_0}.
\label{phase_dyn2}
\end{equation}
We assume that the difference between $T_0$ and $T$ periods is of the order $O(\omega^{-2})$, i.e., $T_0=T+O(\omega^{-2})$. Then we have $(T-T_0)/T_0=(T-T_0)/(T+O(\omega^{-2}))=\Delta[1+O(\omega^{-2})]\approx \Delta$. Finally, we get:
\begin{equation}
M^{\pm}(u_0)= \frac{2\omega^2 \Delta}{\left\langle \Phi^2 \right\rangle I_0^2} .
\label{del_m}
\end{equation}
Since $\Omega=\Omega_0+O(\omega^{-2})$ and $\vartheta^*(t)=t+O(\omega^{-2})$ on the time interval $t\in [0,T]$, we can replace $\Omega$ by $\Omega_0$ and $t$ by $\vartheta^*$  in the Eq.~(\ref{opt_env1}):
\begin{equation}
\psi^*(\Omega_0 \vartheta^*) = \left\lbrace
\begin{array}{l}
1 \; \textrm{when} \; z_{\mathrm{eff}}(\vartheta^*)/u_0>1 \\
0  \; \textrm{when} \; z_{\mathrm{eff}}(\vartheta^*)/u_0<1
\end{array} \right. .
\label{opt_env2}
\end{equation}
This equation is equivalent to the Eq.~(\ref{opt_psi}) of the main text.

Note that the entrainment  is possible only when  $I_0>I_{\textrm{cr}}=\omega\left[2  \Delta/\left\langle \Phi^2 \right\rangle M^{\pm}(0)\right]^{1/2}$. The existence of the critical value $I_{\textrm{cr}}$ is explained as follows. Let's say the frequency mismatch is positive, $\Delta>0$. Then to attain the maximal increase of the phase during the period of oscillations, we have to switch on the perturbation with the maximal amplitude $A=I_0/\omega$ in time intervals where $z_{\mathrm{eff}}(\vartheta(t))$ is positive and switch off the perturbation where  $z_{\mathrm{eff}}(\vartheta(t))<0$ [see Eq. (\ref{phas_dyn_1a})]. Estimating the entrainment threshold with such a stimulation protocol, we define the above critical value $I_{\textrm{cr}}$.

Substituting Eq. (\ref{opt_env2}) into Eq. (\ref{power_lf}), we find the minimal value of the  power functional
\begin{equation}
P_\Omega[A^*\psi^*]= I_0^2 N^{\pm}(u_0)
\label{power_lf1}
\end{equation}
attained with the optimal waveform $A^*\psi^*$. The functions $N^\pm(u)$ are defined in Eq. (\ref{npm}).

\bibliography{references}

\end{document}